# 5G Network Slicing: Analysis of Multiple Machine Learning Classifiers

Mirsad Malkoc
malkocm@sunypoly.edu
State University of New York Polytechnic Institute Utica, NY 13502, UNITED STATES

Hisham A. Kholidy
kholidh@sunypoly.edu
State University of New York Polytechnic Institute Utica, NY 13502, UNITED STATES

*Abstract--* The division of one physical 5G communications infrastructure into several virtual network slices with distinct characteristics such as bandwidth, latency, reliability, security, and service quality is known as 5G network slicing. Each slice is a separate logical network that meets requirements of specific services or use cases, such as virtual reality, gaming, autonomous vehicles, or industrial automation. The network slice can be adjusted dynamically to meet the changing demands of the service, resulting in a more cost-effective and efficient approach to delivering diverse services and applications over a shared infrastructure. This paper assesses various machine learning techniques, including the logistic regression model, linear discriminant model, k-nearest neighbor's model, decision tree model, random forest model, SVC BernoulliNB model, and GaussianNB model, to investigate each model's accuracy and precision on detecting network slices. The report also gives an overview of 5G network slicing.

I. INTRODUCTION

For current cellular communications and the next generation mobile network, meeting high dependability standards, having latency that is low, increased overall capacity, tightened measures for security, and quick connectivity for users is essential [1]. Mobile carriers require a configurable solution to support several tenants on the same infrastructure and next-generation networks support this through Network Slicing (NS), which enables end-to-end allocation of network resources [1]. With more traffic, data-driven decision-making is necessary, and deep learning can improve the performance of 5G networks. In this report, we integrate in-network machine learning and forecasting to construct models, such as algorithms in (ML), to regulate network load efficiency and availability. With the use of important indicators for networking we can assess incoming traffic and forecast the network slice for any unidentified device type using our machine learning approach. We can provide load balancing and effectively utilize the resources of the current network slices thanks to efficient allocation of resources. Even in the case of a network interruption, our machine learning algorithms are capable of making deft decisions and choosing the best network slice. Network slicing will make it possible for many next-generation networks, services, apps, and use cases [1]. Its characteristics enable end-to-end isolation amongst slices while permitting personalization of each slice according to service demands such as capacity, range, security, response time, and dependability [1]. Defending the communications infrastructure system from Distributed Denial of Service (Attacks) entails assuring the separation of assets, data, and network operations between slices [1]. As a result of the increasing and complicated business requirements and expectations of the 5G network, conventional approaches to computer security have proven untenable.

II. Related Works

The article in [2] investigates the most effective methods for next-generation communications slicing as it gains popularity in both the private and public sectors, among the several articles that have been produced on the matter. The article was produced according to studies conducted by professionals from the National security agency and CISA. The paper also presents advice for the best practices for this technology given that communications slicing is a "essential new feature" that allows telecommunications carriers to build multiple virtual networks on top of one physical network. Threats such as DDOS, MITM, and others are explored. The author underlines the necessity of security in the context of next-generation communications slicing, where it is required to separate communications slices in order to guarantee for (CIA) confidentiality, integrity and availability. The report mainly discusses issues including management, performance, and security. According to availability, which relies on the importance of urgent cases and needs of the clients, communication slices are implemented and described in [3]. The article also goes through the fundamental details of three slice kinds that have been standardized by the (eMBB), (IoT), and (URLLC). Next-generation slice committees and that allow us to assess the availability of next-generation slices. The article goes on to illustrate how SDN-based technologies can be used to manage network traffic in order to implement 5G network slicing. The essay also covers network slicing's difficulties and how SDN can facilitate communications slicing at the network layer. The communications network can be divided into several virtual networks using the communication slicing technology that is being suggested, and each virtual network will serve different user types. The report's conclusion claims that availability of communication slices gives the most recent SDN technologies the support they need to help things out. Attempting to map numerous tailored virtualized network requests (also known as services) to a common shared network architecture and allocating network resources to meet various service requirements, the author of [4] presents an effective method for the network slicing problem. The algorithm, while taking into account traffic routing for all services, utilizes an LP dynamic rounding procedure to distribute the virtualized network functions of all services among cloud nodes. The second stage of the method employs an iterative LP refining procedure to find a traffic routing solution for all services that satisfies their end-to-end delay restrictions. The author of [5], explains how network slicing is implemented in a communications network. A method called communications slicing makes it possible to build several virtual networks on top of a single real network infrastructure. In order to improve network dynamics and versatility to accommodate contemporary network applications, the study suggests a network slicing design for a current next-generation network. The previous network architecture must be modified in order to support communications slicing in next-generation networking. Using technology breakthroughs like software-defined networking and network function virtualization, the underlying physical next-generation networking infrastructure is virtualized. After reviewing these related works, we can see that none of these articles apply a machine learning approach on multiple models like I did in my research, or any deep learning in general. They are talking more about the architecture of Next-Generation Communications Slicing, rather than developing models to do a prediction of network slices. The 1st article has a lot of relation to our because, there is talks about denial-of-service attacks just like we talked about it in our report, the author talked about security for communications slicing which is important as well, this is the most related work closest to our report. In [6-35], these previous works are related to the cybersecurity in 5G domain and these works also are related to our machine learning approaches that we used in our paper, they investigate many of the important machine learning models, similar to the ones we used in our paper such as Random Forest, Decision Tree, etc and also deep learning neural networks in these papers, if you want to learn more about cybersecurity in 5G, you should investigate and research these papers, to learn more in depth about 5G networks and machine learning.

III. 5G Network Slicing Architecture

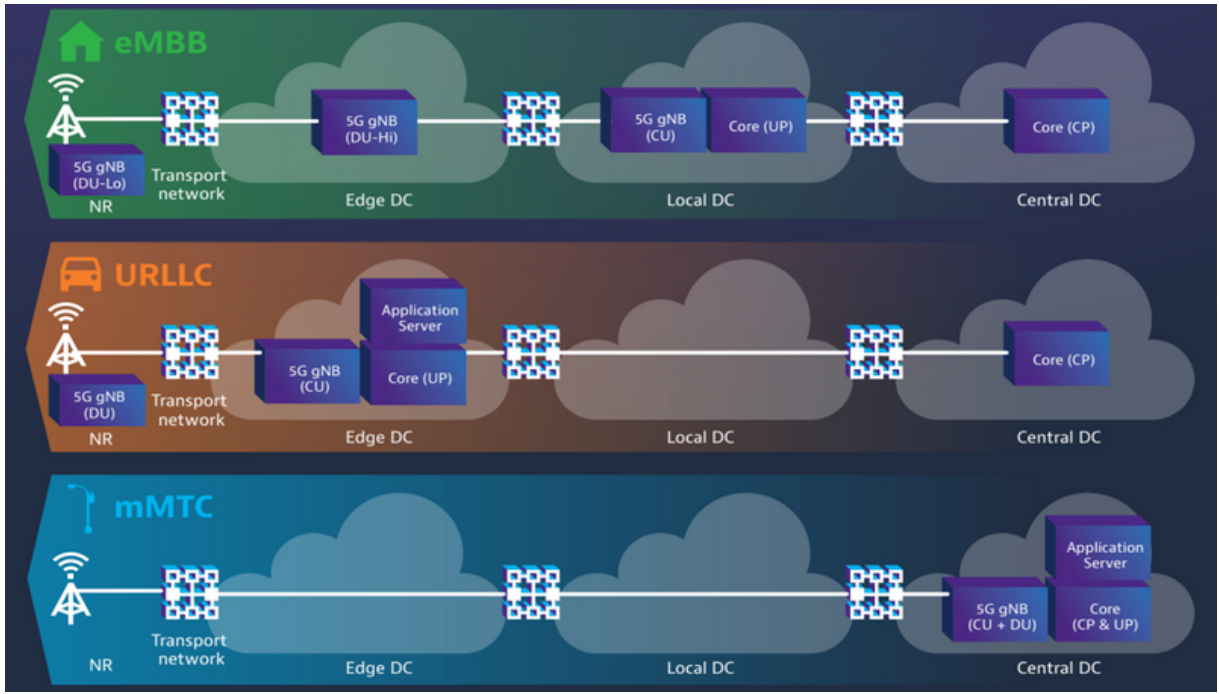

Figure 1 5G Network Slicing Architecture

As shown in Picture 1, in communications slicing, there are the primary categories of services that be given in 5G networks, which include the following, eMBB (enhanced Mobile Broadband, eMBB is intended to provide high-capacity broadband access to smart phones with significant data transfer rates, reduced latency, and high dependability [36]. This type of service is suitable for applications such as streaming video, online gaming, and other data-intensive applications. Network slices for eMBB typically allocate a large amount of bandwidth and low latency to meet the high data transfer requirements. uRLLC (ultra-Reliable Low-Latency Communications), uRLLC is designed to support crucial business applications which need reduced latencies or high dependability, such as industrial automation, remote surgery, and autonomous vehicles [36]. The network slice for uRLLC is designed to deliver shorter transmission delay, have higher reliability, and has to have short packet error rate for meeting the stringent requirements of these applications. mMTC (massive Machine-Type Communications) mMTC supports large-scale communication between machines, sensors, and other devices [36]. This type of service requires low power consumption, low cost, and massive connectivity. Network slices for mMTC are designed to provide many connections per unit area, support low-cost devices, and optimize power consumption. Each service has unique bandwidth, latency, reliability, and connectivity requirements. Network slicing allows operators to allocate resources dynamically based on the specific needs of each service, enabling them to optimize network performance and ensure that services meet their particular requirements. Network Slicing enables operators to provide services catering to different use cases and applications, enhancing the overall user experience [36].

IV. 5G Network Slicing Layers and Use Cases

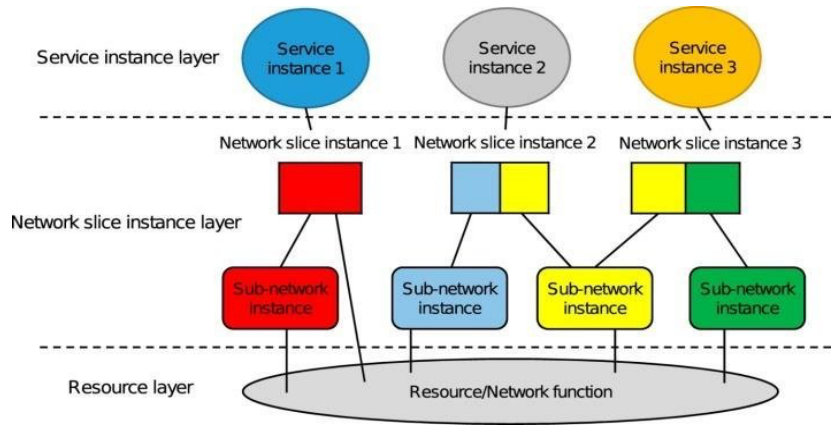

Figure 2 Network Slicing Layers

This idea of Communications Slicing, as explained in reference [37], is made of three primary layers, all having a unique purpose. The layers are:

Service instance layer: As shown in Figure 2, The above layer reflects the various services that need to be provided so each application is expressed by a instance of a service. Such solutions can be offered by the provider or even a related party [37].

Communications slicing instance layer: As shown in Figure 2, A communications slice occurrence is created by a communications slice plan and so it specifies the communications properties that can be shared across various different instances. It can be built up of neither, 1 or maybe more subnetwork instances which may be replicated through other communications instances [37].

Service level: The leadership of the network slice is carried out at this layer, as shown in Figure 2 [37].

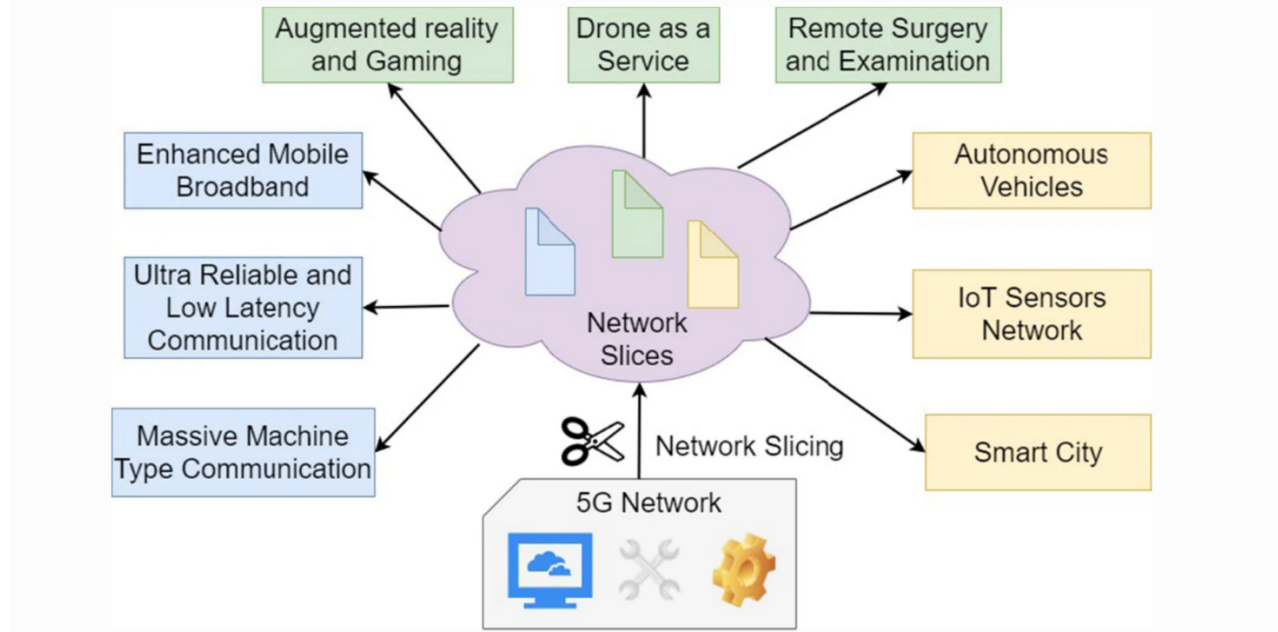

Figure 3 Network Slicing Use Cases

Looking at Figure 3, the 5G network has to provide a broad range of application cases, which require different communications slices like shown by Figure 3, the 5G network must service a broad variety of applications, which require diverse network slices tailored to their unique demands and requirements [37]. The emergence of IoT has considerably broadened the applications that the 5G network must accommodate. To fulfill these various objectives, communications slicing must be implemented flexibly and dynamically which can adjust to the shifting complexity of use case needs. Different application cases have various properties, such as quick throughput, extremely low latency, latency that is steady and reduced, large speed, and massive IoT application cases. Mega IoT constitutes one of the most expected application cases for the next-generation communications network, which relies on slicing mobile 5G communications to connect integrated sensors all around the world. This application case covers cities that are connected, asset management, intelligent grids, and farming sectors [37]. Super, reduced pings use cases utilize high uptime and connectivity for remote control of key assets, power system control, automatization of enterprises, robotic systems, and unmanned aerial vehicles.

Improved performance levels application cases involve quality of VR videos to be increased, increased multimedia, as well as strong quality experiences of multimedia. Picture 3 depicts an essential representation of various application cases for the communication slices of 5G has to accommodate [37]. The different slices are color-coded to indicate their varying requirements, with similar slices grouped. Each use case necessitates a unique network slice, and the created slices cater to multiple use cases. Integrating use cases into the classification is crucial as they help differentiate the types of slices needed. The 5G network must support numerous use cases, each with distinct network service requirements and slice characteristics. By leveraging the slicing algorithm, numerous slicing could get generated as well as distributed for such certain application cases. Thus, incorporating application cases for classification was essential in identifying the nature of slices the 5G network must create. The classification encompasses various subfactors under use cases, including portability, management systems, protection, home automation, reduced latencies, and extreme throughput application cases, to cater for demands for industries, companies, and clients [37]. However, mostly applications, such as portability, pose a significant challenge as they have to service various elements of client portability. Along with reality, portability necessitates flexibility or dynamism in the communications slicing serving the clients portability requirement. Thankfully, communications slicing enables the design of an extremely flexible interconnection. [37].

V. Importance of 3GPP Network Slicing

The 3GPP has acknowledged the significance of communications slicing for the fifth generation internetworking landscape. As such, it has become a significant focus for working groups developing the 5G core architecture. This recognition has been reflected in the technical specifications of the 3GPP, with TS 23.501 defining stage 2 to include network slicing and TS 22.261 outlining the necessary provisions for creating, associating devices with, and ensuring performance isolation during both normal and elastic slice operation. The 3GPP's Release 16 of the 5G specification has introduced several opportunities, including low-latency industrial IoT and autonomous driving. As part of this release, the 5G core solutions have been developed to facilitate cellular IoT [37]. The implications of using an unlicensed NR spectrum regarding bandwidth and cost have been evaluated. [37]

VI. Challenges Network Slicing Faces

There are many challenges that network slicing faces, such as Resource Sharing, NGMN have identified that the sharing of resources among different communications slicing neighbors is a critical concern. Sharing of resources can be achieved through fixed segment sharing and elastic share that is dynamic [38][39]. However, because of the rapidly changing nature of network load, dynamic resource sharing can be more efficient regarding resource utilization. Resource sharing presents specific challenges that need to be addressed. For instance, telecommunication assets may be shared among RAN segments, requiring some robust rf timing algorithm for distributing spectrum resources between these segments. [38][39]

Moreover, other types of resource sharing, such as computational data exchange, should be considered. While data exchange may benefit network operators, and moreover confronts issues including segment isolation. Another challenge that communications slicing faces is that implementing network slicing heavily relies on virtualization technology, which has become the most crucial technology for this purpose. Virtualization technology has developed significantly in the last two decades, mainly used in wired networks, with core network virtualization receiving much attention. However, wireless communication is more complicated due to the uncertainty of variable time in nature for connections that are wireless and their susceptibility to interference. Several virtualization methods used in wired networks really aren't directly relevant to wire-free communications. Therefore, developing new virtualization mechanisms that enable distribution access radio spectrum assets and virtualizing transmitters is crucial for implementing RAN slicing [38][39]. Another challenge that network slicing faces is Security; ensuring data integrity for communications slicing is a challenging issue that must be properly handled, primarily since resources are shared against many different communications segments handling various services, each with additional security policy requirements. Thus, designing security protocols for network slicing must examine the influence for certain slices as well as the entire infrastructure system [38][39].

Furthermore, implementing network slicing in a multi-domain infrastructure makes security issues even more complex, necessitating the design of defense policy coordinating techniques across various disciplines [38][39]. The network slicing technology enables communication operators to quickly construct highly scalable deployments, allowing various applications to also have unique rational slicing implementations on something like a public infrastructure. This, alongside other solutions like MEC, SDN, or NFV, will become an cornerstone as the service-oriented next-generation connectivity especially concerning the implementation of network slicing over 5G networks [38][39].

VII. 5G Radio Access Networks, SDN

The Heterogeneous Wireless Network is a significant element of next-generation communications and is being tested in telecommunications sector from the emergence of cellular technology. Over time, RAN became evolved across several generations of mobile connectivity, through 1G and 2G through 3G, 4G, and finally, 5G. [40] Overall, primary components of RAN is both the transmitters and panels, which provide network coverage globally. The radio sites within RAN provide radio access, coordinating and managing different resources across multiple radio sites. RAN transmits signals to various wireless endpoints worldwide and other network traffic. Several types of RANS, including GRAN, GERAN, UTRAN, and E-UTRAN, each provides different services [40]. The RAN controller plays a crucial role in controlling the connected nodes and providing functions like control of radio spectrum, vehicular networks, and encryption techniques. The current Radio Access Network designs are primarily built based separating management but also client planes, and RAN controllers exchange user messages through SDN switches. Such division fits to NFV and SDN approaches, which could be connected with the 5G network design enabling communications slicing. Consequently, NFV is a critical component of 5G architecture and plays a vital role in network slicing. [40] NFV offers several benefits, such as, eliminating the need for specialized hardware, allowing service providers to move network functions to cloud-based servers, including routers and firewalls, Placing network functions closer to the radio interface or in the cloud, leading to a more agile and flexible network that can meet the demands of network traffic in real-time, optimizing the performance of different applications, even those with varying bandwidth, latency, availability, and security requirements, increasing monitoring and logging options, providing better visibility into the system and enhancing the ability to identify anomalies or prevent vulnerability exploitations[40].

VIII. Machine Learning with Kaggle Network Slicing Dataset Using Colab

In this section, we will be running Machine Learning Code using Python using the Google Colab

Environment; you can use the steps that will be provided below to run code similar to mine to solve any machine learning problem you may have in Cybersecurity by interpreting these results you will be able to understand which Machine Learning Model is best for each situation that you may face, whether it would be email spam or any other cybersecurity attacks. Using these Machine Learning models, we can predict which network slice is the best for each use case. The dataset that we used came from Kaggle. It was one of the top network slicing datasets on Google, which is why we picked it; the other datasets are not preprocessed and not easy to use like this dataset; if you are looking for datasets, you should always look for and find a preprocessed dataset. This is the meaning of each item in our dataset, LTE/5g - User Equipment categories or classes are utilized to specify the performance standards. Packet Loss Rate is the number of communications never delivered divided by an overall number of transmissions transmitted. Transmission Latency refers to the time for a transmission to be received. A Slice type is a network arrangement that allows for the operation of multiple networks that are virtualized and self-contained. GBR refers to Guaranteed Bit Rate, which is a certain level of network bandwidth that is assured. The utilization of LTE/5G technology in the Healthcare sector is represented by the value 1 or 0. Industry 4.0's application in digital businesses is represented by the value 1 or 0. IoT Devices' application is represented by their usage. The usage of LTE/5G technology in Public Safety, for the purpose of public welfare and safety, is represented by the value 1 or 0. Smart City & Home represents the usage of this technology in day-to-day household chores. Smart Transportation is the usage of this technology in public transportation. The usage of smartphones for cellular data transmission is indicated.

Figure 4 Network Slicing Training Dataset

In Figure 4, you can see the Network Slicing Training Dataset, which contains 31,584 rows. Some columns have LTE/5G, Time, Packet Loss, Packet Delay, IoT, GBR, Non-GBRM, AR/VR/Gaming, Healthcare, Industry, IoT Devices, Public Safety, Smart City, Smart Transportation, Smartphone, Slice Type.

Picture 5 Communications Slicing Test Dataset

Looking at, Picture 5, you can see the Network Slicing Test Dataset, which contains 31,585 rows. Some columns have LTE/5G, Time, Packet Loss, Packet Delay, IoT, GBR, Non-GBRM, AR/VR/Gaming, Healthcare, Industry, IoT Devices, Public Safety, Smart City, Smart Transportation, Smartphone. This new data can be used to predict what network slice is best for each use case scenario.

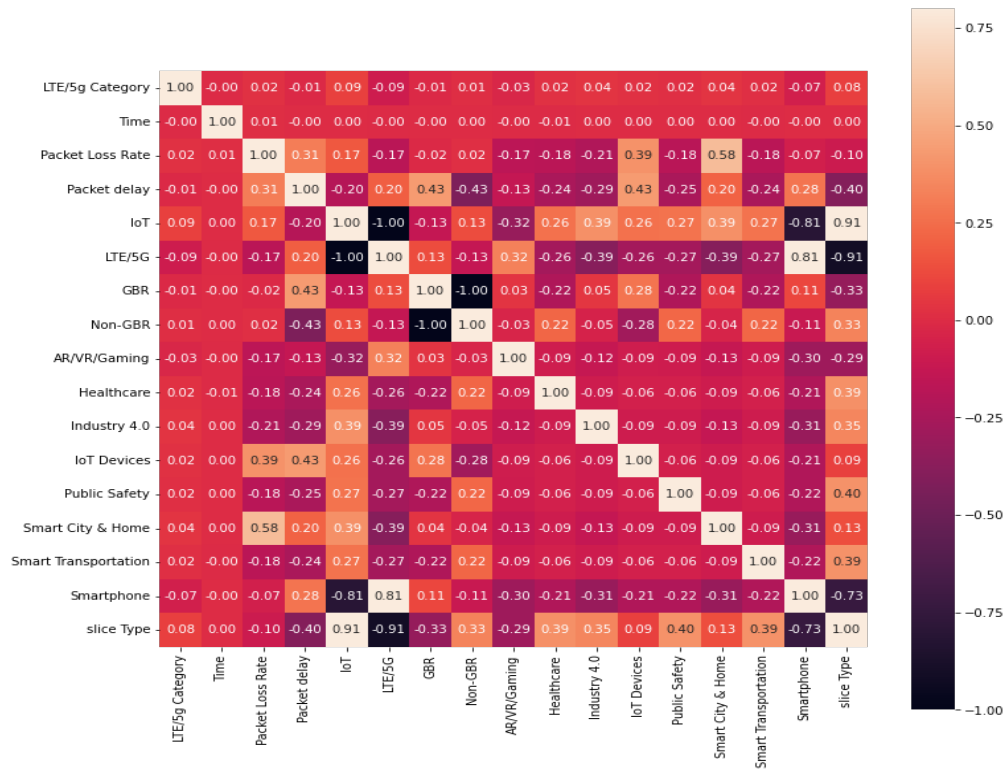

Figure 6 Heatmap of 5G Network Slicing Dataset

In Figure 6, we generated the Heatmap of the 5G Network Slicing Dataset by using some simple python code; this heatmap is used for creating a correlation between the data in the dataset and the features; this is an excellent tool that can be used for any machine learning problem, it is located in the seaborn library which can be easily imported by using a simple command just as shown in the steps below, and you run a couple of lines of python code to generate this heatmap.

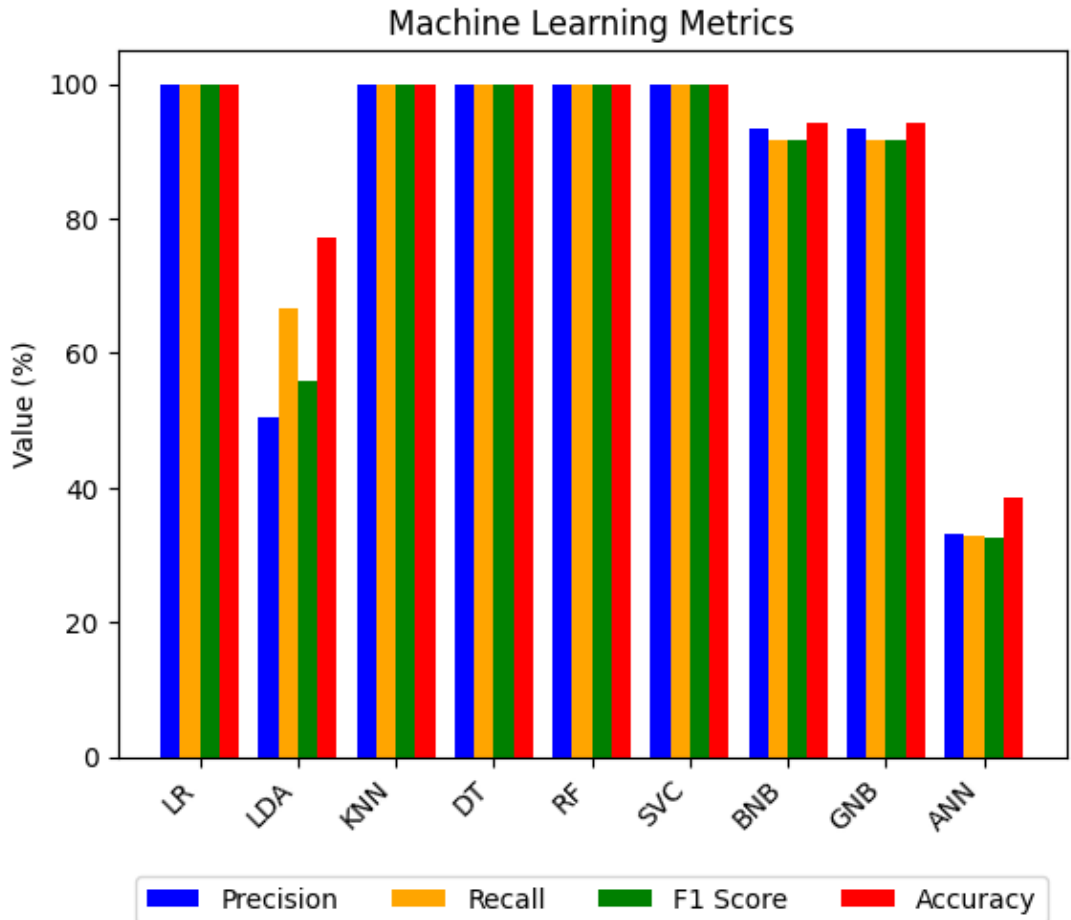

Figure 7 Machine Learning Metrics Analysis

In Figure 7, we compared the precision, recall, f1 score, and accuracy, between all of our models, we were able to see that LR, KNN, DT, RF, SVC all gave a 100% score for precision, recall, f1 score, and accuracy, these are examples of perfect models to use for a machine learning problem like this. Looking at LDA, BNB, and GNB, we were able to see that the scores were much less than 100%, LDA had a 50.29% precision, recall of 66.67%, f1 score of 55.81%, and a 77.16% accuracy, this is a very bad model to use to choose network slices, also BNB, and GNB, gave us the same results in terms of precision, recall, f1 score, and accuracy, the precision was 93.43%, the recall was 91.52%, the f1 score was 91.50%, and the accuracy was 94.19%. Overall BNB, and GNB are both good models to use for this type of problem but they are not the best, LDA should be avoided due to low precision, recall, f1 score, and low accuracy. Looking at ANN, the neural network, it received the lowest scores, the precision was 33.00%, the recall was 32.77%, f1 score was 32.57%, and the accuracy was 38.53%. ANN and LDA should be completely avoided for this specific type of network slicing problem that we have solved.

Here is the what these metrics mean:

Precision: The ratio of true positives (positive occurrences that were accurately predicted) to all positive examples is known as precision. It gauges how accurate or reliable the model's optimistic forecasts are. A high precision means that there aren't many false positives and that the model is highly good at predicting positive events [41].

Recall: Recall is the ratio of the number of actual positive cases to the number of genuine positives. It gauges how thorough the model's positive predictions are. A high recall means that there are minimal false negatives and that the model is able to catch the majority of real positive events [41].

F1 Score: A single indicator of the model's accuracy is the F1 score, which is the harmonic mean of precision and recall. It is the weighted average of recall and precision, with weights based on beta values (usually set to 1). A high F1 score suggests a precision to recall ratio that is well balanced [41].

Accuracy: Accuracy is the ratio of correctly predicted events to all predicted events. It evaluates how accurately the model's forecasts are made in general. For datasets that are imbalanced is when one class has a significantly higher number of instances than the other, so accuracy might not be the best metric.

```python
import pandas as pd
import numpy as np
import matplotlib.pyplot as plt
import seaborn as sns
from scipy.stats import skew, kurtosis
import matplotlib.pyplot as plt
from sklearn import svm, datasets
from sklearn.model_selection import train_test_split
from sklearn.metrics import plot_roc_curve, auc
```

```python
data = pd.read_csv('/content/train_dataset.csv')
data.head()
```

| | LTE/5g Category | Time | Packet Loss Rate | Packet delay | IoT | LTE/5G | GBR | Non-GBR | AR/VR/Gaming | Healthcare | Industry 4.0 | IoT Devices | Public Safety | Smart City & Home | Smart Transportation | Smartphone | slice Type |
|---|---|---|---|---|---|---|---|---|---|---|---|---|---|---|---|---|---|
| 0 | 14 | 0 | 0.000001 | 10 | 1 | 0 | 0 | 1 | 0 | 0 | 0 | 0 | 1 | 0 | 0 | 0 | 3 |
| 1 | 18 | 20 | 0.001000 | 100 | 0 | 1 | 1 | 0 | 1 | 0 | 0 | 0 | 0 | 0 | 0 | 0 | 1 |
| 2 | 17 | 14 | 0.000001 | 300 | 0 | 1 | 0 | 1 | 0 | 0 | 0 | 0 | 0 | 0 | 0 | 1 | 1 |
| 3 | 3 | 17 | 0.010000 | 100 | 0 | 1 | 0 | 1 | 0 | 0 | 0 | 0 | 0 | 0 | 0 | 1 | 1 |
| 4 | 9 | 4 | 0.010000 | 50 | 1 | 0 | 0 | 1 | 0 | 0 | 0 | 0 | 0 | 1 | 0 | 0 | 2 |

```python
data.shape
```
(31583, 17)

Figure 8 Dataset Contents First Five Columns

In Figure 8, we imported the necessary libraries for our machine learning dataset and looked at the dataset contents for the first five columns using the data.head() command, and printed the shape of our dataset by using the data—shape command.

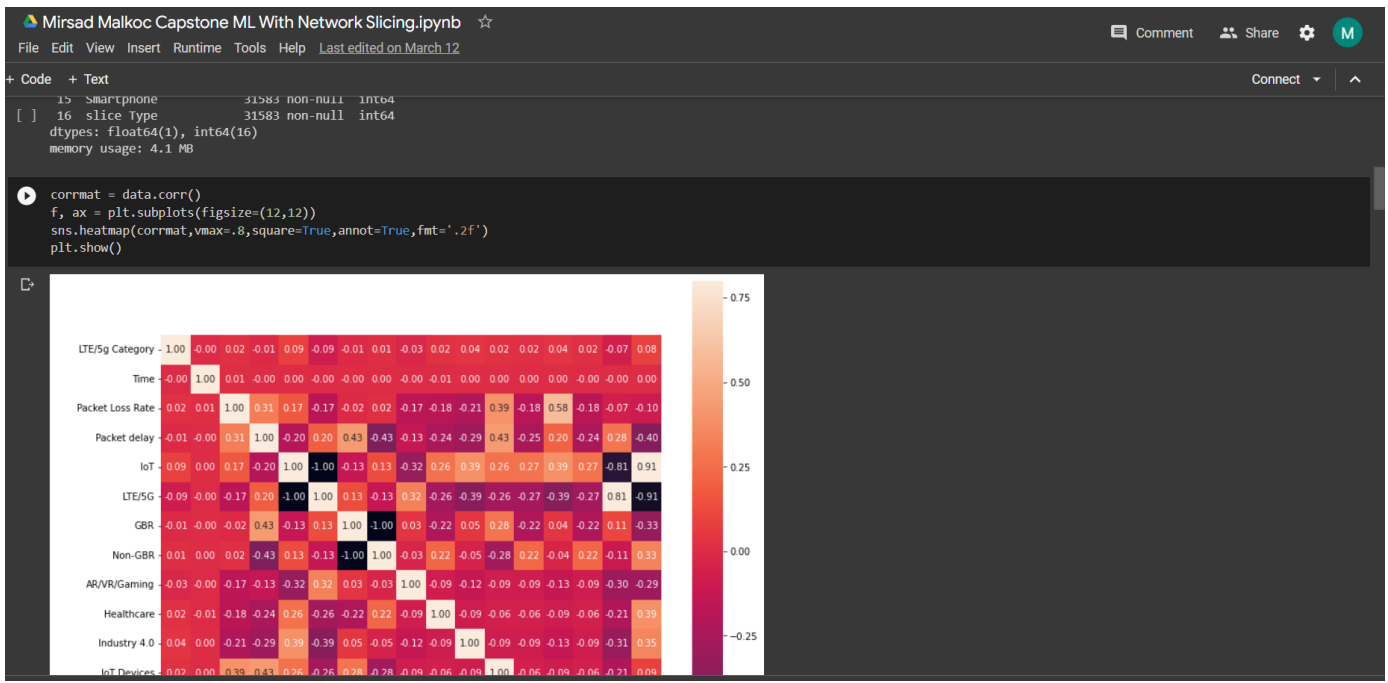

Figure 9 All Columns of Dataset

In Figure 9, we printed all of the columns of our dataset using the data.info() command; this can be used for any dataset; we had 16 columns in the dataset.

Figure 10 Creating Heatmap

In Figure 10, we created a Heatmap using simple code using the Seaborn library that is included with Colab; you can do the same with any dataset if you want to

```python
from sklearn.model_selection import train_test_split
```

```python
X_train, X_test, y_train, y_test = train_test_split(X,y,test_size=0.2,random_state=10)
```

```python
print(X_train.shape)
print(X_test.shape)
print(y_train.shape)
print(y_test.shape)
```

```
(25266, 7)
(6317, 7)
(25266,)
(6317,)
```

```python
from sklearn.discriminant_analysis import LinearDiscriminantAnalysis
from sklearn.linear_model import LogisticRegression
from sklearn.neighbors import KNeighborsClassifier
from sklearn.tree import DecisionTreeClassifier
from sklearn.ensemble import RandomForestClassifier
from sklearn.naive_bayes import BernoulliNB, GaussianNB
from sklearn.svm import SVC
from sklearn.model_selection import KFold, StratifiedKFold, cross_val_score
from sklearn.metrics import accuracy_score, confusion_matrix, classification_report, roc_auc_score, roc_curve
from sklearn.preprocessing import StandardScaler
```

```python
sc = StandardScaler()
X_train = sc.fit_transform(X_train)
X_test = sc.transform(X_test)
```

Figure 11 Training All Models

In Figure 11, we used the train_test_split to divide our datasets into training as well as testing datasets; then I used a test_size of 0.2 and a random state of 10 then we printed the shapes of by x and y for both training and testing, then we imported all of the classifiers that we are using for training our dataset on these models. StandardScaler is used to standardize the data values into a standard format.

```python
models = []
models.append(('LR',LogisticRegression(solver='liblinear',multi_class='ovr')))
models.append(('LDA',LinearDiscriminantAnalysis(solver='svd')))
models.append(('KNN',KNeighborsClassifier(n_neighbors=10,metric='minkowski')))
models.append(('DT',DecisionTreeClassifier(criterion='gini')))
models.append(('RF',RandomForestClassifier(n_estimators=200,criterion='gini',max_depth=None)))
models.append(('SVM',SVC(C=1.0,kernel='rbf',degree=3,gamma='auto')))
models.append(('BNB',BernoulliNB()))
models.append(('GNB',GaussianNB()))
```

```python
cvresults = []
names = []
res = []
for name, model in models:
    skf = StratifiedKFold(n_splits=10)
    cvres = cross_val_score(model,X_train,y_train,cv=skf,scoring='accuracy')
    cvresults.append(cvres)
    names.append(name)
    res.append(cvres.mean())
    print("Name:",name," Mean cross val res:",cvres.mean())
plt.ylim(.900, .999)
plt.bar(names, res, color ='maroon', width = 0.6)

plt.title('Comparison Between Algorithms')
plt.show()
```

```
Name: LR  Mean cross val res: 1.0
Name: LDA  Mean cross val res: 0.7653368481837173
Name: KNN  Mean cross val res: 1.0
Name: DT  Mean cross val res: 1.0
Name: RF  Mean cross val res: 1.0
Name: SVM  Mean cross val res: 1.0
Name: BNB  Mean cross val res: 0.9424522990185803
Name: GNB  Mean cross val res: 0.9424522990185803
```

Figure 12 Mean Cross Val Of All Models

In Figure 12, we printed to the screen all of the models that we trained and the Mean Cross Val Res for each model, the LR, KNN, DT, RF, and SVM, had the highest scores, these models should be used for a network slicing prediction they are all good to use and we should avoid using the rest for this certain situation.

[Screenshot of Jupyter notebook showing Logistic Regression classifier code and output]

```
LR = LogisticRegression(solver='liblinear',multi_class='ovr')
LR.fit(X_train, y_train)
y_pred = LR.predict(X_test)
accuracy_score(y_test,y_pred)
```

1.0

```
confusion_matrix(y_test,y_pred)
```

array([[3380,    0,    0],
       [   0, 1494,    0],
       [   0,    0, 1443]])

```
print(classification_report(y_test,y_pred))
```

|  | precision | recall | f1-score | support |
|---|---|---|---|---|
| 1 | 1.00 | 1.00 | 1.00 | 3380 |
| 2 | 1.00 | 1.00 | 1.00 | 1494 |
| 3 | 1.00 | 1.00 | 1.00 | 1443 |
| accuracy |  |  | 1.00 | 6317 |
| macro avg | 1.00 | 1.00 | 1.00 | 6317 |
| weighted avg | 1.00 | 1.00 | 1.00 | 6317 |

Figure 13 Logistic Regression Classifier

In Figure 13, we trained the Logistic Regression Classifier on the 5G Network Slicing dataset; using the above commands, we printed the confusion matrix and the classification report to the screen; we viewed the results and analyzed them. We found that the Logistic Regression has achieved an accuracy of 100%; this is one of the highest scores for the models we have trained our dataset on. Looking at the confusion matrix, we were able to tell that class 1 had a 100% recall, class 2 had a 100% recall, and class 3 had a 100% recall, adding up all of the numbers in the confusion matrix we received a total of 6317 true positives, as shown on the classification report as well

[Screenshot of Jupyter notebook showing Linear Discriminant Analysis code and output]

|  | precision | recall | f1-score | support |
|---|---|---|---|---|
| 2 | 1.00 | 1.00 | 1.00 | 1494 |
| 3 | 1.00 | 1.00 | 1.00 | 1443 |
| accuracy |  |  | 1.00 | 6317 |
| macro avg | 1.00 | 1.00 | 1.00 | 6317 |
| weighted avg | 1.00 | 1.00 | 1.00 | 6317 |

```
LDA = LinearDiscriminantAnalysis(solver='svd')
LDA.fit(X_train, y_train)
y_pred = LDA.predict(X_test)
accuracy_score(y_test,y_pred)
```

0.7634953300617382

```
confusion_matrix(y_test,y_pred)
```

array([[3380,    0,    0],
       [   0,    0, 1494],
       [   0,    0, 1443]])

```
print(classification_report(y_test,y_pred))
```

|  | precision | recall | f1-score | support |
|---|---|---|---|---|
| 1 | 1.00 | 1.00 | 1.00 | 3380 |
| 2 | 0.00 | 0.00 | 0.00 | 1494 |
| 3 | 0.49 | 1.00 | 0.66 | 1443 |
| accuracy |  |  | 0.76 | 6317 |
| macro avg | 0.50 | 0.67 | 0.55 | 6317 |
| weighted avg | 0.65 | 0.76 | 0.69 | 6317 |

Figure 14 Linear Discriminant Analysis

In Figure 14, we trained the Linear Discriminant Analysis Classifier on the 5G Network Slicing dataset; using the commands shown above, we printed the confusion matrix screen and the classification report to the screen; we viewed the results and analyzed them, and we found that the Linear Discriminant Analysis Classifier has achieved an accuracy of 76.35%, this is one of the lowest scores for the models that we have trained our dataset on. Looking at the confusion matrix, we were able to tell that class 1 had a 100% recall, class 2 had a no recall because of having no data, and class 3 had a 49.132% recall, adding up all of the numbers in the confusion matrix we received a total of 6317 true positives, as shown on the classification report as well

Figure 15 K-Nearest Neighbors Classifier

In Figure 15, we trained the K-Nearest Neighbors Classifier on the 5G Network Slicing dataset; using the above commands, we printed the confusion matrix and the classification report to the screen; we viewed the results and analyzed them. We found that the K-Nearest Neighbors Classifier has achieved an accuracy of 100%; this is one of the highest scores for the models we have trained our dataset on. Looking at the confusion matrix, we were able to tell that class 1 had a 100% recall, class 2 had a 100% recall, and class 3 had a 100% recall, adding up all of the numbers in the confusion matrix we received a total of 6317 true positives, as shown on the classification report as well

Figure 16 Decision Tree Classification Model

In Figure 16, we trained the Decision Tree Classification Model on the 5G Network Slicing dataset using the commands shown above; we printed the confusion matrix to the screen and the classification report to the screen, we viewed the results and analyzed them, and we found that the Decision Tree Classifier has achieved an accuracy of 100%, this is one of the highest scores for the models that we have trained our dataset on. Looking at the confusion matrix, we were able to tell that class 1 had a 100% recall, class 2 had a 100% recall, and class 3 had a 100% recall, adding up all of the numbers in the confusion matrix we received a total of 6317 true positives, as shown on the classification report as well

Figure 17 Random Forest Classifier

In Figure 17, we trained the Random Forest Classifier on the 5G Network Slicing dataset using the commands shown above; we printed the confusion matrix to the screen and the classification report to the screen, we viewed the results and analyzed them, and we found that the Random Forest Classifier has achieved an accuracy of 100%, this is one of the highest scores for the models that we have trained our dataset on. Looking at the confusion matrix, we were able to tell that class 1 had a 100% recall, class 2 had a 100% recall, and class 3 had a 100% recall, adding up all of the numbers in the confusion matrix we received a total of 6317 true positives, as shown on the classification report as well

Figure 18 SVC Classifier

In Figure 18, we trained the Support Vector Machine Classifier on the 5G Network Slicing dataset using the commands shown above; we printed the confusion matrix to the screen and the classification report to the screen, we viewed the results and analyzed them, and we found that the Support Vector Machine Classifier has achieved an accuracy of 100%, this is one of the highest scores for the models that we have trained our dataset on. Looking at the confusion matrix, we were able to tell that class 1 had a 100% recall, class 2 had a 100% recall, and class 3 had a 100% recall, adding up all of the numbers in the confusion matrix we received a total of 6317 true positives, as shown on the classification report as well

```
              precision    recall  f1-score   support

           1       1.00      1.00      1.00      3380
           2       1.00      1.00      1.00      1494
           3       1.00      1.00      1.00      1443

    accuracy                           1.00      6317
   macro avg       1.00      1.00      1.00      6317
weighted avg       1.00      1.00      1.00      6317
```

```python
BNB= BernoulliNB()
BNB.fit(X_train, y_train)
y_pred = BNB.predict(X_test)
accuracy_score(y_test,y_pred)
```

0.9419028019629571

```python
confusion_matrix(y_test,y_pred)
```

```
array([[3380,    0,    0],
       [   0, 1494,    0],
       [   0,  367, 1076]])
```

```python
print(classification_report(y_test,y_pred))
```

```
              precision    recall  f1-score   support

           1       1.00      1.00      1.00      3380
           2       0.80      1.00      0.89      1494
           3       1.00      0.75      0.85      1443

    accuracy                           0.94      6317
   macro avg       0.93      0.92      0.91      6317
weighted avg       0.95      0.94      0.94      6317
```

Figure 19 Bernoulli Naive Bayes Classifier

In Figure 19, we trained the Bernoulli Naïve Bayes Classifier on the 5G Network Slicing dataset using the commands shown above; we printed the confusion matrix onto the screen and the classification report to the screen, we viewed the results and analyzed them, and we found that the Bernoulli Naive Bayes Classifier has achieved an accuracy of 94%, this is a high score but still not as good as the other classifiers we looked at. Looking at the confusion matrix, we were able to tell that class 1 had a 100% recall, class 2 had an 80.279% recall, and class 3 had a 100% recall; adding up all of the numbers in the confusion matrix, we received a total of 6317 true positives, as shown on the classification report as well.

```python
GNB= GaussianNB()
GNB.fit(X_train, y_train)
y_pred = GNB.predict(X_test)
accuracy_score(y_test,y_pred)
```

0.9419028019629571

```python
confusion_matrix(y_test,y_pred)
```

```
array([[3380,    0,    0],
       [   0, 1494,    0],
       [   0,  367, 1076]])
```

```python
print(classification_report(y_test,y_pred))
```

```
              precision    recall  f1-score   support

           1       1.00      1.00      1.00      3380
           2       0.80      1.00      0.89      1494
           3       1.00      0.75      0.85      1443

    accuracy                           0.94      6317
   macro avg       0.93      0.92      0.91      6317
weighted avg       0.95      0.94      0.94      6317
```

Figure 20 Gaussian Naive Bayes Classifier

In Figure 20, we trained the Gaussian Naïve Bayes Classifier on the 5G Network Slicing dataset; using the above commands, we printed the confusion matrix and the classification report to the screen; we viewed the results and analyzed them. We found that the Gaussian Naive Bayes Classifier has achieved an accuracy of 94%; this is a high score but still not as good as the other classifiers we looked at. Looking at the confusion matrix, we were able to tell that class 1 had a 100% recall, class 2 had an 80.279% recall, and class 3 had a 100% recall; adding up all of the numbers in the confusion matrix we received a total of 6317 true positives, as shown on the classification report as well.

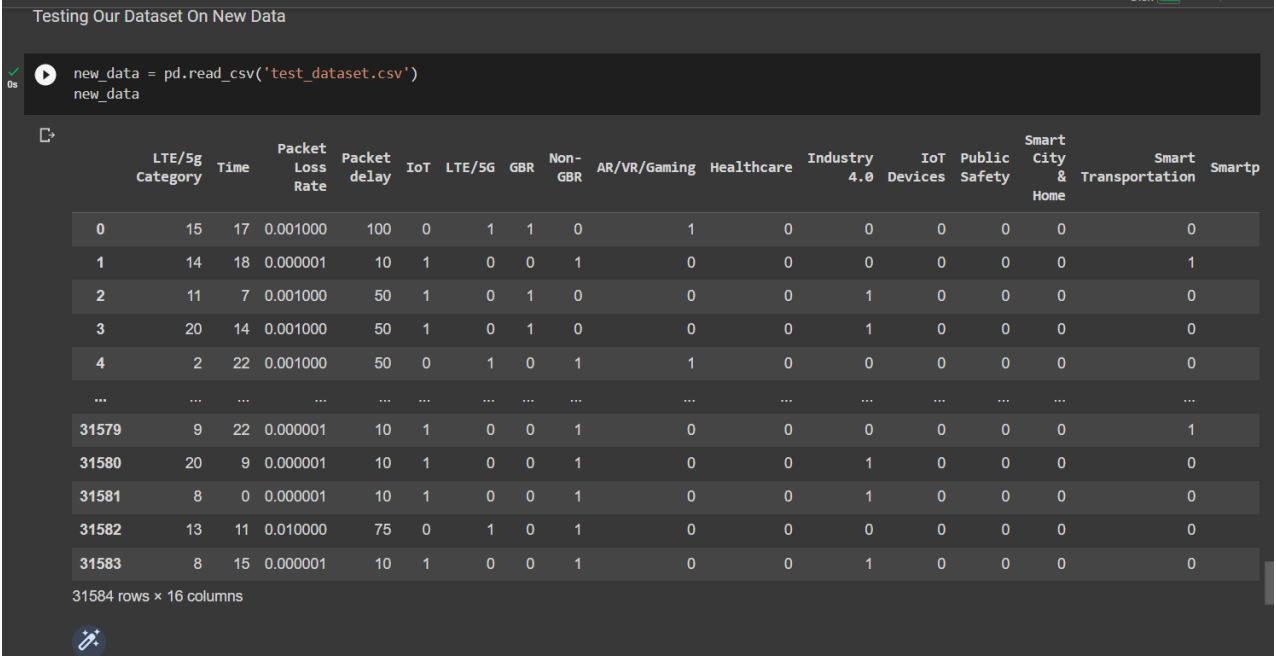

Figure 21 Importing the test dataset to be used for prediction with model

In Figure 21, we imported the test dataset so that we can use it for predictions, to determine what network slice is best for each use case.

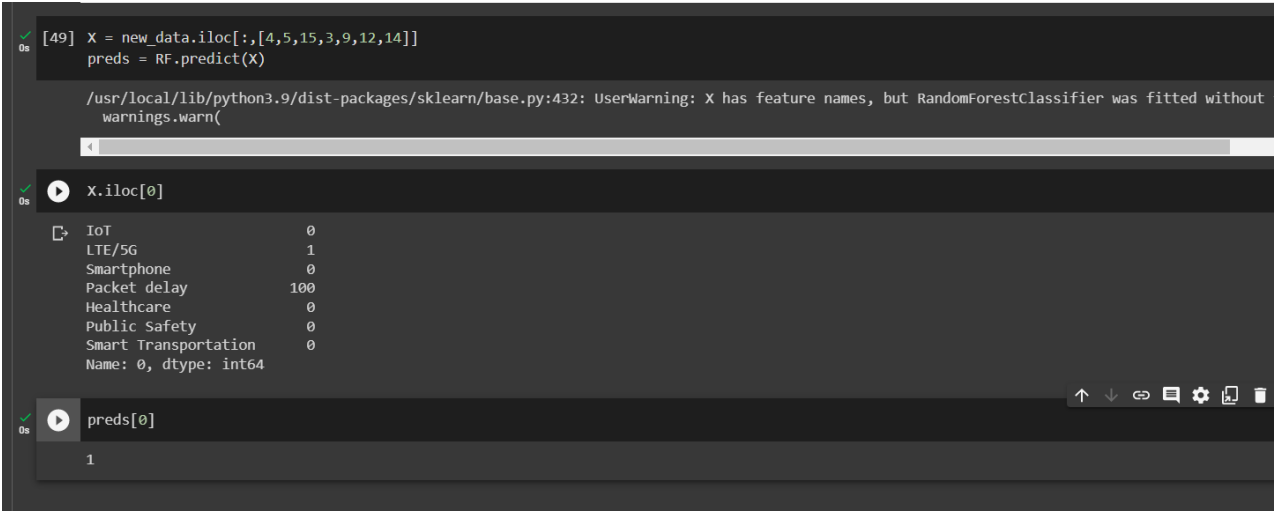

Figure 22 Predicting a network slice using RF Classifier

```
[124] X = new_data.iloc[:,[4,5,15,3,9,12,14]]
      preds = RF.predict(X)

      /usr/local/lib/python3.9/dist-packages/sklearn/base.py:432: UserWarning: X has feature names, but RandomForestClassifier was fitted without feature names
        warnings.warn(

[125] X.iloc[6]

      IoT                    1
      LTE/5G                 0
      Smartphone             0
      Packet delay          50
      Healthcare             0
      Public Safety          0
      Smart Transportation   0
      Name: 6, dtype: int64

[126] preds[6]

      2
```

Figure 23 Predicting another network slice using RF Classifier

In Figure 22, we used columns 4,5,15,3,9,12,14 for prediction this is the requirement for all of our classifiers, you only need 7 features to test on which is 7 of the columns, it can be any of them, and then we used the Random Forest Classifier as it had 100% percent accuracy, and we were able to predict that it belonged to network slice 1 this was determined based on the given usage and delay in packet delivery to destination node. In Figure 23, we did also prediction on row 6 for the same columns and we came up with slice 2 which was accurate according to our dataset. Random Forest is a very good classifier for this problem, however you can use any model that we trained to predict the network slice type.

```python
#Importing necessary Libraries
import numpy as np
import pandas as pd
import tensorflow as tf

[153] #Initialising ANN
ann = tf.keras.models.Sequential()
#Adding First Hidden Layer
ann.add(tf.keras.layers.Dense(units=64,activation="relu"))
#Adding Second Hidden Layer
ann.add(tf.keras.layers.Dense(units=64,activation="relu"))
#Adding Output Layer
ann.add(tf.keras.layers.Dense(units=1,activation="sigmoid"))

[154] #Compiling ANN
ann.compile(optimizer="adam",loss="binary_crossentropy",metrics=['accuracy'])

[155] #Fitting ANN
ann.fit(X_train,y_train,batch_size=32,epochs = 100)

Epoch 1/100
790/790 [==============================] - 2s 2ms/step - loss: -5995.9077 - accuracy: 0.5273
Epoch 2/100
790/790 [==============================] - 1s 2ms/step - loss: -92600.3125 - accuracy: 0.5311
Epoch 3/100
790/790 [==============================] - 1s 2ms/step - loss: -369352.5625 - accuracy: 0.5311
Epoch 4/100
790/790 [==============================] - 1s 2ms/step - loss: -888919.0000 - accuracy: 0.5311
Epoch 5/100
790/790 [==============================] - 1s 2ms/step - loss: -1674171.6250 - accuracy: 0.5311
Epoch 6/100
790/790 [==============================] - 1s 2ms/step - loss: -2760189.5000 - accuracy: 0.5311
Epoch 7/100
790/790 [==============================] - 2s 2ms/step - loss: -4163048.0000 - accuracy: 0.5311
Epoch 8/100
790/790 [==============================] - 2s 3ms/step - loss: -5914207.5000 - accuracy: 0.5311
Epoch 9/100
790/790 [==============================] - 1s 2ms/step - loss: -8033875.0000 - accuracy: 0.5311
Epoch 10/100
790/790 [==============================] - 1s 2ms/step - loss: -10540706.0000 - accuracy: 0.5311
Epoch 11/100
```

Figure 24 ANN (Artificial Neural Network)

```
[155] Epoch 86/100
2m   790/790 [==============================] - 2s 2ms/step - loss: -3174668800.0000 - accuracy: 0.5311
     Epoch 87/100
     790/790 [==============================] - 1s 2ms/step - loss: -3278808576.0000 - accuracy: 0.5311
     Epoch 88/100
     790/790 [==============================] - 1s 2ms/step - loss: -3385120768.0000 - accuracy: 0.5311
     Epoch 89/100
     790/790 [==============================] - 2s 2ms/step - loss: -3493756672.0000 - accuracy: 0.5311
     Epoch 90/100
     790/790 [==============================] - 2s 2ms/step - loss: -3604488192.0000 - accuracy: 0.5311
     Epoch 91/100
     790/790 [==============================] - 1s 2ms/step - loss: -3717542400.0000 - accuracy: 0.5311
     Epoch 92/100
     790/790 [==============================] - 1s 2ms/step - loss: -3833093888.0000 - accuracy: 0.5311
     Epoch 93/100
     790/790 [==============================] - 1s 2ms/step - loss: -3950900224.0000 - accuracy: 0.5311
     Epoch 94/100
     790/790 [==============================] - 1s 2ms/step - loss: -4070866688.0000 - accuracy: 0.5311
     Epoch 95/100
     790/790 [==============================] - 1s 2ms/step - loss: -4193505792.0000 - accuracy: 0.5311
     Epoch 96/100
     790/790 [==============================] - 1s 2ms/step - loss: -4318584832.0000 - accuracy: 0.5311
     Epoch 97/100
     790/790 [==============================] - 2s 2ms/step - loss: -4445705728.0000 - accuracy: 0.5311
     Epoch 98/100
     790/790 [==============================] - 2s 3ms/step - loss: -4575473664.0000 - accuracy: 0.5311
     Epoch 99/100
     790/790 [==============================] - 1s 2ms/step - loss: -4707803648.0000 - accuracy: 0.5311
     Epoch 100/100
     790/790 [==============================] - 1s 2ms/step - loss: -4842585600.0000 - accuracy: 0.5311
     <keras.callbacks.History at 0x7fd63873ab20>
```

Figure 25 ANN (Artificial Neural Network Results)

In Figure 24 and Figure 25, we tried a basic Artificial Neural Network, to test deep learning on this dataset, the results weren't too good even after changing around parameters, since the dataset isn't too big and neural networks require larger datasets, we ended up with a 53% accuracy for a neural network which isn't good we used 64 layers for the first hidden layer, and 64 layers for the second hidden layer and 1 for the output, we tried it on 100 epochs and it still wasn't enough for this dataset, we can see this by the loss that is stated it is in the negatives, machine learning seems to be best suited for this dataset.

IX. Conclusion and Discussion about Machine Learning with Colab

Communications slicing constitutes one of the fundamental elements of next-generation wireless networking and is expected to play a crucial role in defining the development of next generation wireless networks. Communications slicing enables the communications providers to develop virtualized connections in a single physical network infrastructure, enabling them can deliver tailored connectivity services that can be tailored for specific demands of varied usage application cases. Communications operators could create a network slice optimized for quick response time with high throughput for applications like autonomous vehicles or remote surgery, while another slice may be optimized for high capacity and coverage for use cases such as streaming video or IoT devices. By enabling multiple virtual networks to coexist on a single physical network, network slicing can help operators optimize network resources and improve network efficiency, while delivering a great deal of adaptability that accommodate varying customer needs. This makes network slicing a key enabler for 5G use cases that require specific network characteristics and performance, such as industrial IoT, smart cities, and augmented reality.

I was delighted with working with Google Collaborate to implement this code; it came out as expected, just like my previous machine learning experience, we came out with good results on the models, and we were able to determine what models were best, the only thing that we were not able to figure out, in this case, was the ROC curves, which cannot be displayed for a multi-class classification problem like this one, We can use the results above to determine the ROC curve, so it isn't needed to understand what is going on in the machine learning training process. It is incredible to find such a good dataset that yielded almost 100% accuracy for many of the models we tested; this is rare, especially the datasets that you need to run preprocessing on; you would probably never be able to get 100% accuracy and recall. With the correct code, it didn't take Google Colab a lot of time to train these models that we tested; it took probably about 2-3 mins to train our dataset on these models, using the Google Collaborate Environment. At the end of this report, you should have been able to learn more about Network Slicing and Machine Learning in Cybersecurity; these are two crucial things you must remember for the future.

## X. Comparison of my ML Results to others

After looking around, we weren't able to find any 5G Network Slicing articles or papers, that used a similar approach to mine, however looking at source [11], where we found our dataset, we found some code that was ran by the author of the dataset, we looked at their results and they took a much larger approach than us, but they automated their machine learning models so they didn't have to write any code, the feature is known as LazyPredict, the code is located at [19], here are their results:

t[15]:

| Model | Accuracy | Balanced Accuracy | ROC AUC | F1 Score | Time Taken |
|---|---|---|---|---|---|
| AdaBoostClassifier | 1.00 | 1.00 | None | 1.00 | 0.94 |
| BaggingClassifier | 1.00 | 1.00 | None | 1.00 | 0.17 |
| XGBClassifier | 1.00 | 1.00 | None | 1.00 | 1.24 |
| SVC | 1.00 | 1.00 | None | 1.00 | 0.24 |
| SGDClassifier | 1.00 | 1.00 | None | 1.00 | 0.16 |
| RidgeClassifierCV | 1.00 | 1.00 | None | 1.00 | 0.16 |
| RidgeClassifier | 1.00 | 1.00 | None | 1.00 | 0.07 |
| RandomForestClassifier | 1.00 | 1.00 | None | 1.00 | 0.91 |
| Perceptron | 1.00 | 1.00 | None | 1.00 | 0.14 |
| PassiveAggressiveClassifier | 1.00 | 1.00 | None | 1.00 | 0.10 |
| NuSVC | 1.00 | 1.00 | None | 1.00 | 40.71 |
| NearestCentroid | 1.00 | 1.00 | None | 1.00 | 0.08 |
| LogisticRegression | 1.00 | 1.00 | None | 1.00 | 0.81 |
| LinearSVC | 1.00 | 1.00 | None | 1.00 | 2.39 |
| LabelSpreading | 1.00 | 1.00 | None | 1.00 | 29.94 |
| LabelPropagation | 1.00 | 1.00 | None | 1.00 | 14.73 |
| KNeighborsClassifier | 1.00 | 1.00 | None | 1.00 | 3.45 |
| GaussianNB | 1.00 | 1.00 | None | 1.00 | 0.04 |
| ExtraTreesClassifier | 1.00 | 1.00 | None | 1.00 | 0.79 |
| ExtraTreeClassifier | 1.00 | 1.00 | None | 1.00 | 0.04 |
| DecisionTreeClassifier | 1.00 | 1.00 | None | 1.00 | 0.07 |
| CalibratedClassifierCV | 1.00 | 1.00 | None | 1.00 | 9.19 |
| BernoulliNB | 1.00 | 1.00 | None | 1.00 | 0.07 |
| LGBMClassifier | 1.00 | 1.00 | None | 1.00 | 0.53 |
| LinearDiscriminantAnalysis | 0.87 | 0.83 | None | 0.87 | 0.21 |
| QuadraticDiscriminantAnalysis | 0.23 | 0.33 | None | 0.09 | 0.10 |
| DummyClassifier | 0.39 | 0.33 | None | 0.39 | 0.04 |

Figure 26 Overall ML Models Results

```
                        SVC

              precision    recall  f1-score   support

           1       1.00      1.00      1.00      3360
           2       1.00      1.00      1.00      1479
           3       1.00      1.00      1.00      1478

    accuracy                           1.00      6317
   macro avg       1.00      1.00      1.00      6317
weighted avg       1.00      1.00      1.00      6317
```

Figure 27 SVC Model

```
        RandomForestClassifier

              precision    recall  f1-score   support

           1       1.00      1.00      1.00      3360
           2       1.00      1.00      1.00      1479
           3       1.00      1.00      1.00      1478

    accuracy                           1.00      6317
   macro avg       1.00      1.00      1.00      6317
weighted avg       1.00      1.00      1.00      6317
```

Figure 28 RF Classifier

```
        LogisticRegression

              precision    recall  f1-score   support

           1       1.00      1.00      1.00      3360
           2       1.00      1.00      1.00      1479
           3       1.00      1.00      1.00      1478

    accuracy                           1.00      6317
   macro avg       1.00      1.00      1.00      6317
weighted avg       1.00      1.00      1.00      6317
```

Figure 29 LR Classifier

```
        LinearDiscriminantAnalysis

              precision    recall  f1-score   support

           1       0.82      0.97      0.89      3360
           2       1.00      0.75      0.86      1479
           3       0.92      0.76      0.83      1478

    accuracy                           0.87      6317
   macro avg       0.91      0.83      0.86      6317
weighted avg       0.88      0.87      0.87      6317
```

Figure 30 LDA Model

```
              KNeighborsClassifier

              precision    recall  f1-score   support

           1       1.00      1.00      1.00      3360
           2       1.00      1.00      1.00      1479
           3       1.00      1.00      1.00      1478

    accuracy                           1.00      6317
   macro avg       1.00      1.00      1.00      6317
weighted avg       1.00      1.00      1.00      6317
```

Figure 31 KNN Model

```
              GaussianNB

              precision    recall  f1-score   support

           1       1.00      1.00      1.00      3360
           2       1.00      1.00      1.00      1479
           3       1.00      1.00      1.00      1478

    accuracy                           1.00      6317
   macro avg       1.00      1.00      1.00      6317
weighted avg       1.00      1.00      1.00      6317
```

Figure 32 GNB Model

```
              DecisionTreeClassifier

              precision    recall  f1-score   support

           1       1.00      1.00      1.00      3360
           2       1.00      1.00      1.00      1479
           3       1.00      1.00      1.00      1478

    accuracy                           1.00      6317
   macro avg       1.00      1.00      1.00      6317
weighted avg       1.00      1.00      1.00      6317
```

Figure 33 DT Model

```
                     BernoulliNB

              precision    recall  f1-score   support

           1       1.00      1.00      1.00      3360
           2       1.00      1.00      1.00      1479
           3       1.00      1.00      1.00      1478

    accuracy                           1.00      6317
   macro avg       1.00      1.00      1.00      6317
weighted avg       1.00      1.00      1.00      6317
```

Figure 34 BNB Model

Looking at Figures 26-34 we can compare their results to my results for ML model classification, it seems like many of their classifiers scored a 100% accuracy for both columns, several like ours like the BernoulliNB, GaussianNB DecisionTreeClassifier, K-Nearest-Neighbors, Random Forest Classifier, SVC, Logistic Regression, all received scores of 100% accuracy, and F1 scores of 100%, Linear Discriminant Analysis received a score of 87% accuracy, which was more than ours which came out to 77% accuracy, the LazyPredict most likely did a better fine tuning of the code that was used to train the model. Overall, my Machine Learning analysis was very similar to this approach except that I coded my own models, they used LazyPredict to generate theirs without any code. If you aren't good at making code, you can look into LazyPredict where you have to write minimal code and it will train all the models for you. Here are the ROC Curves for all the classification models we used in our report,

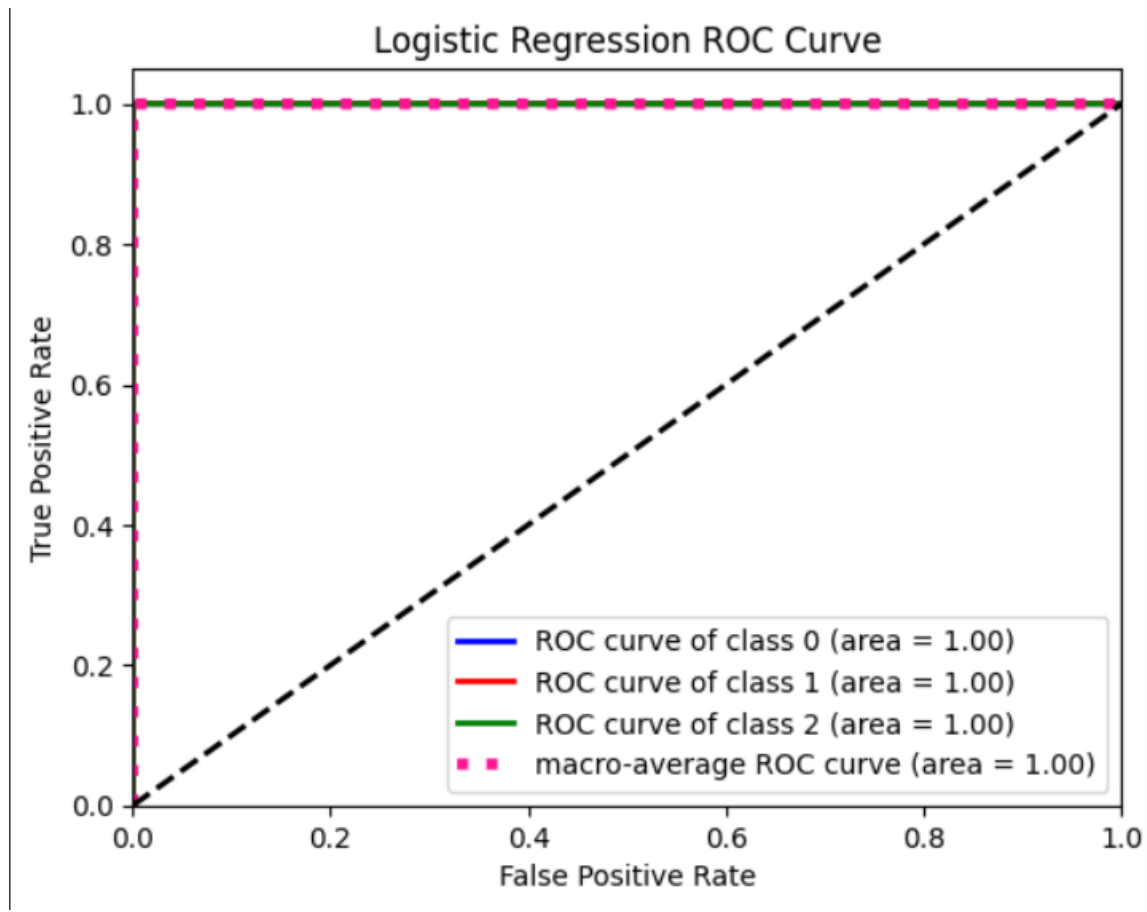

Figure 35 Logistic Regression ROC Curve

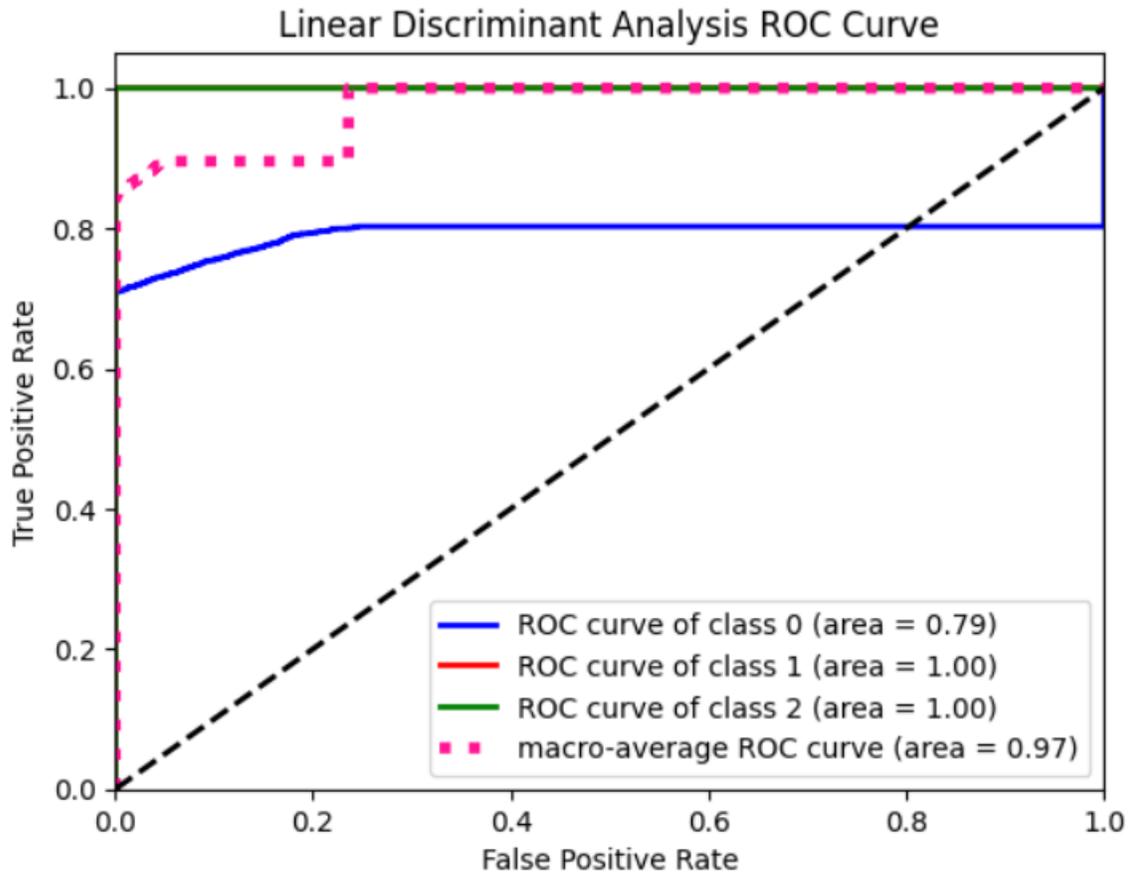

Figure 36 Linear Discriminant Analysis ROC Curve

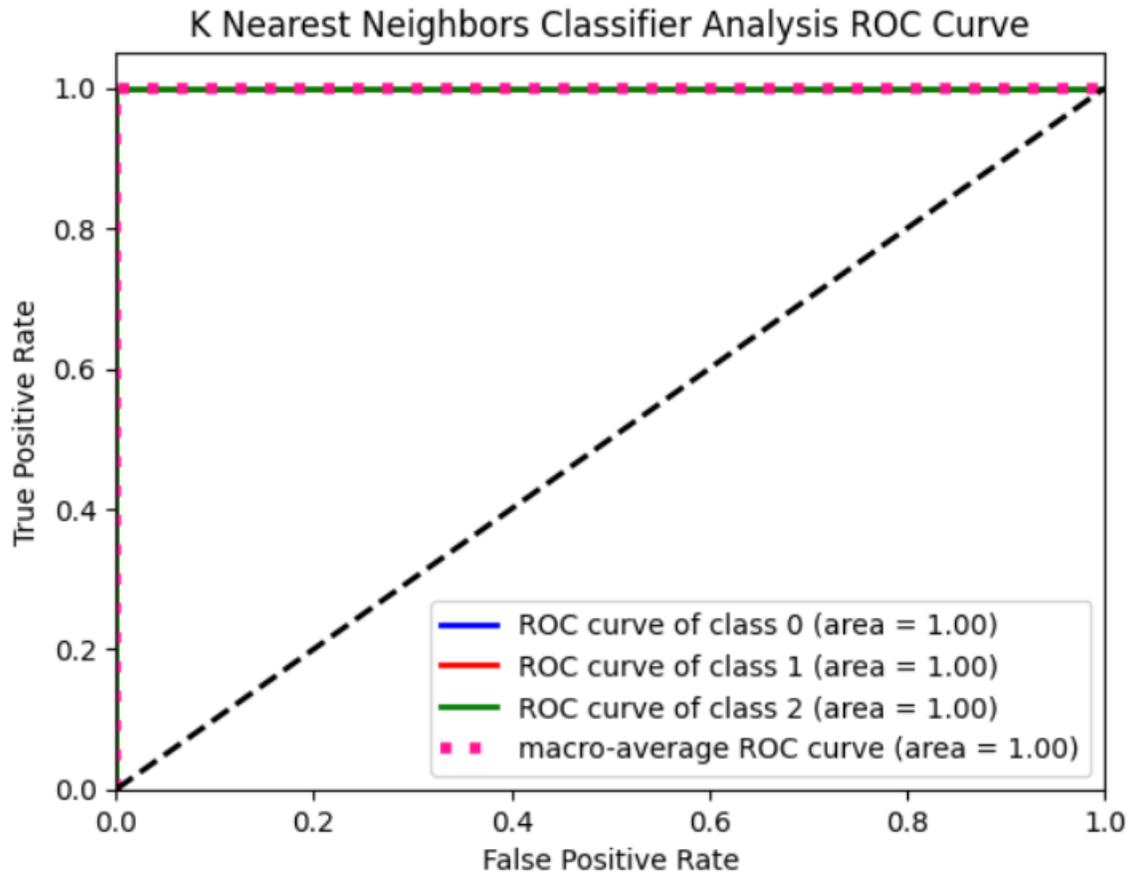

Figure 37 K-Nearest Neighbors ROC Curve

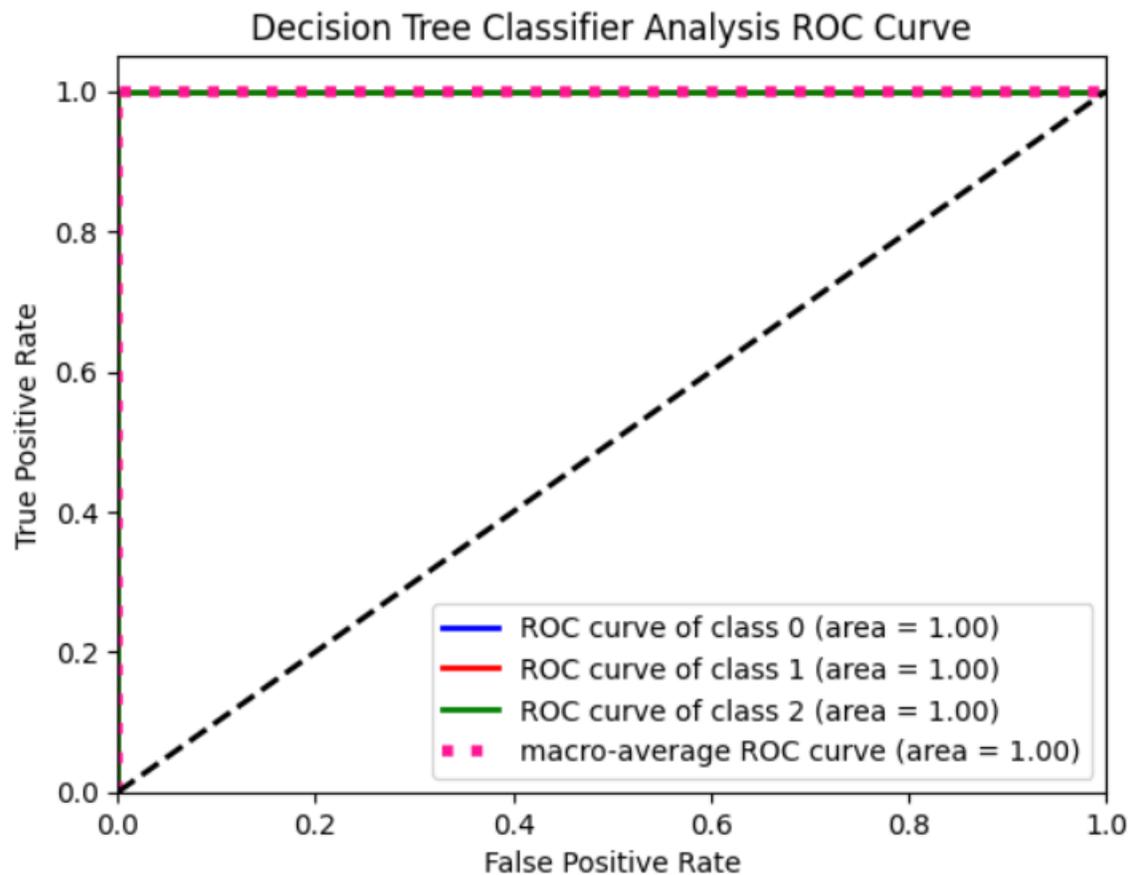

Figure 38 Decision Tree ROC Curve

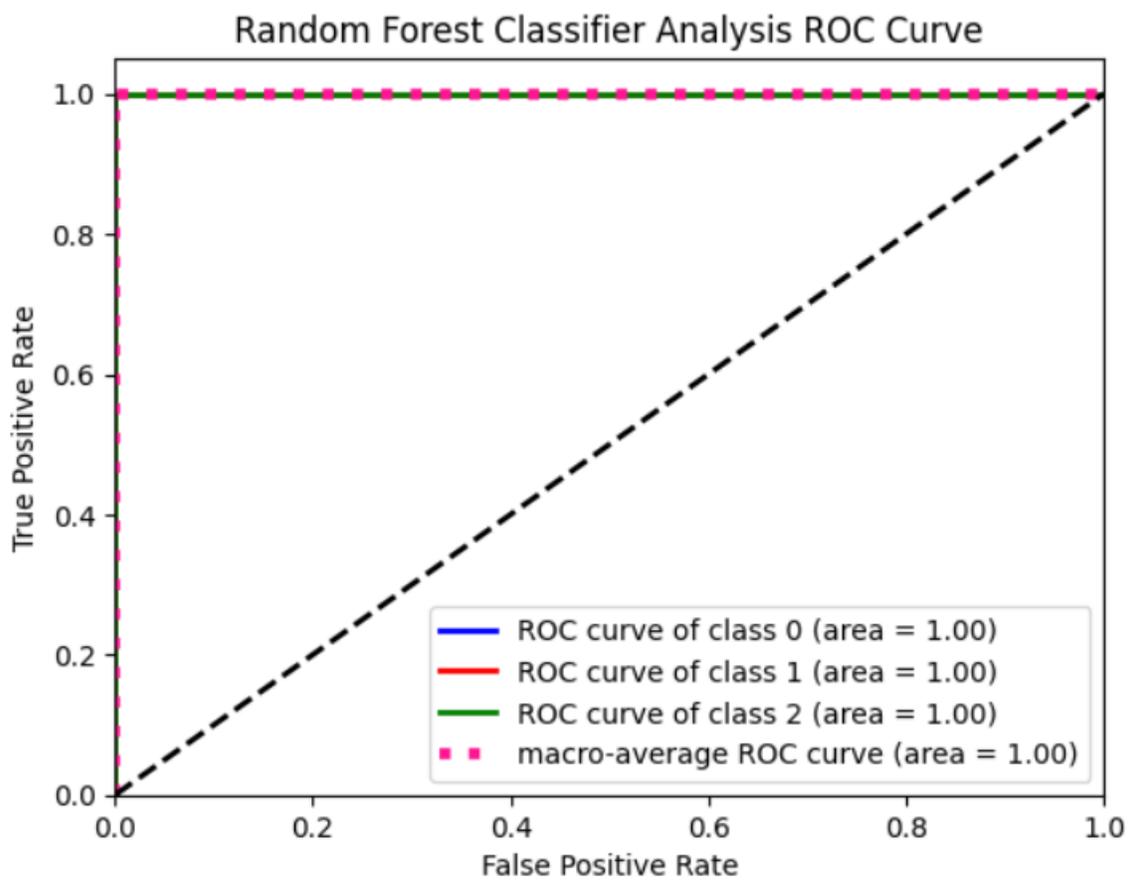

Figure 39 Random Forest ROC Curve

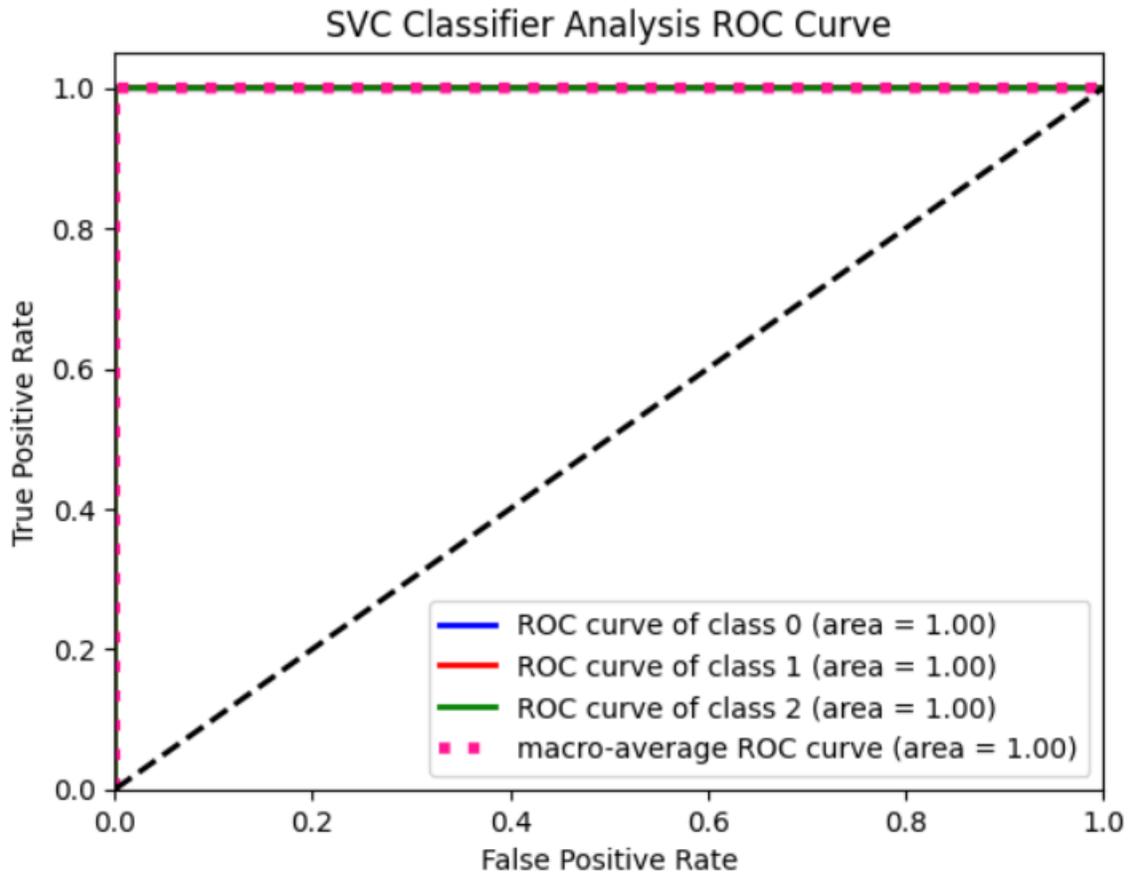

Figure 40 SVC ROC Curve

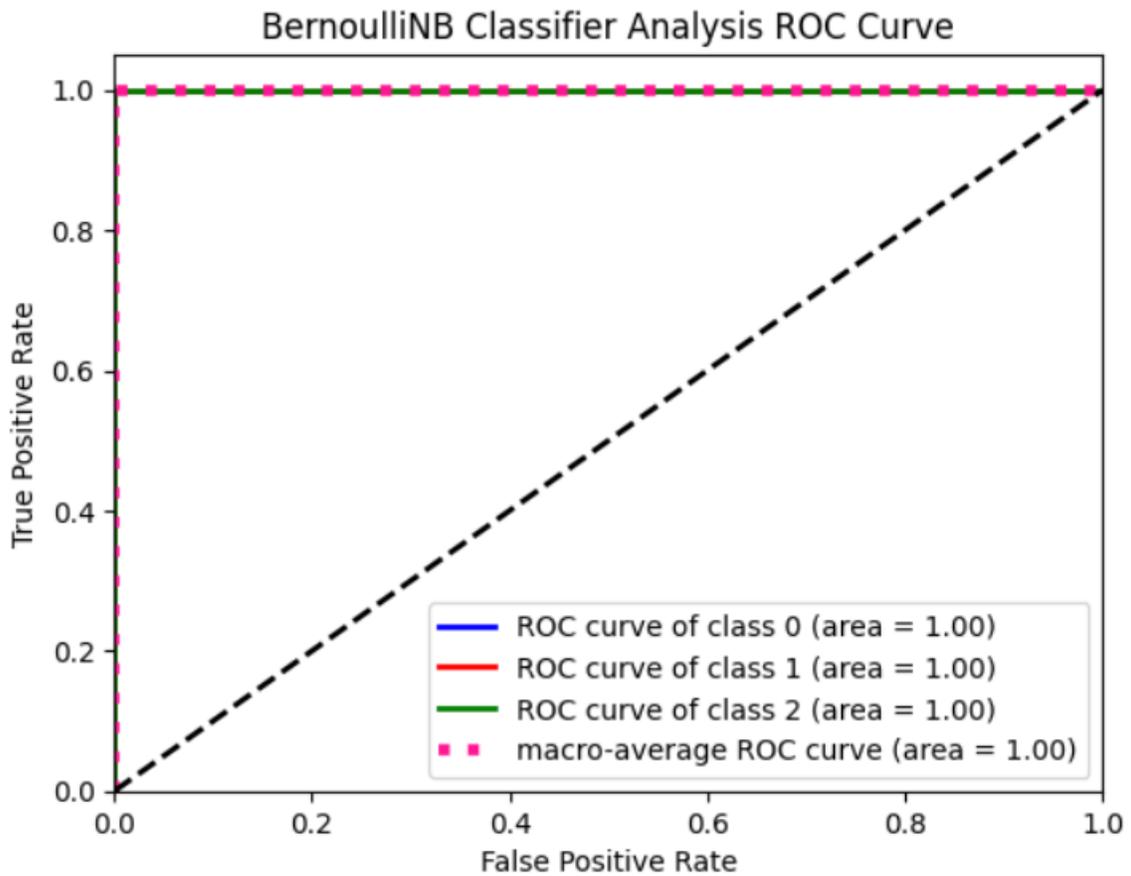

Figure 41 BernoulliNB ROC Curve

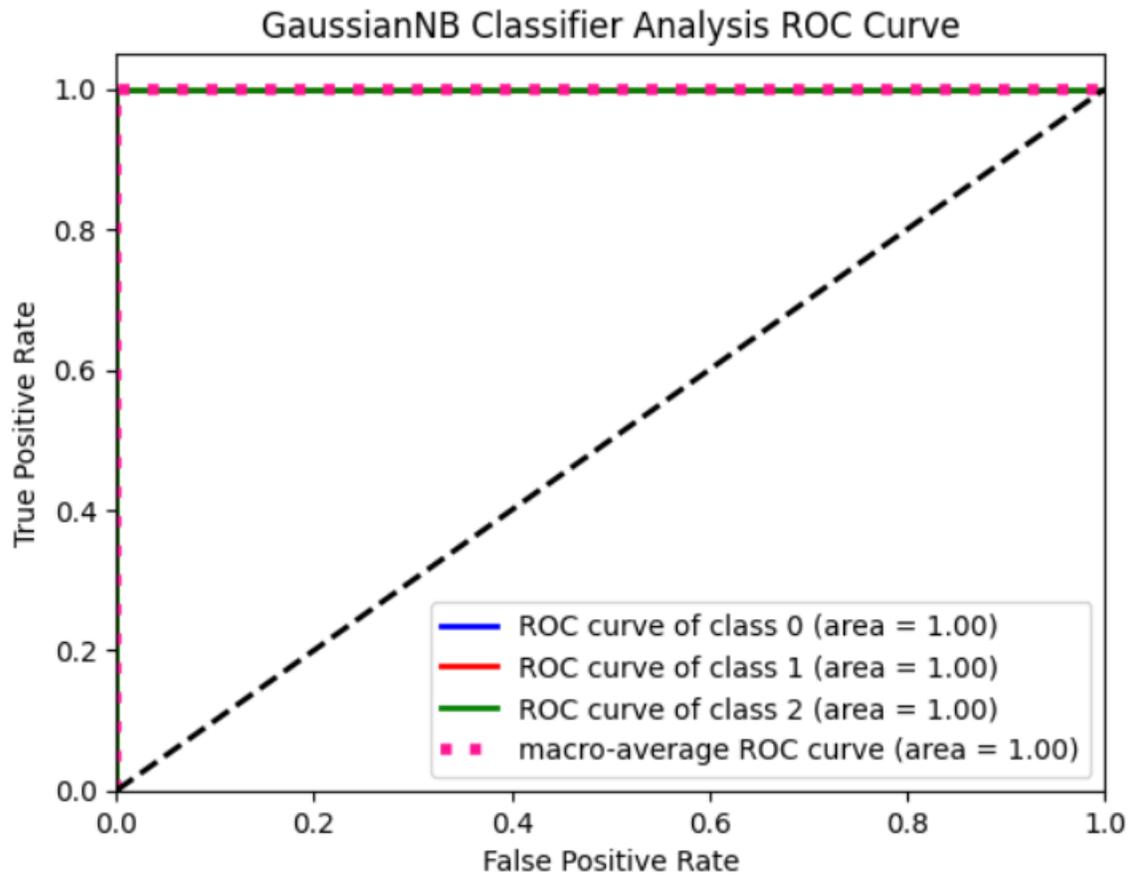

Figure 42 GaussianNB ROC Curve

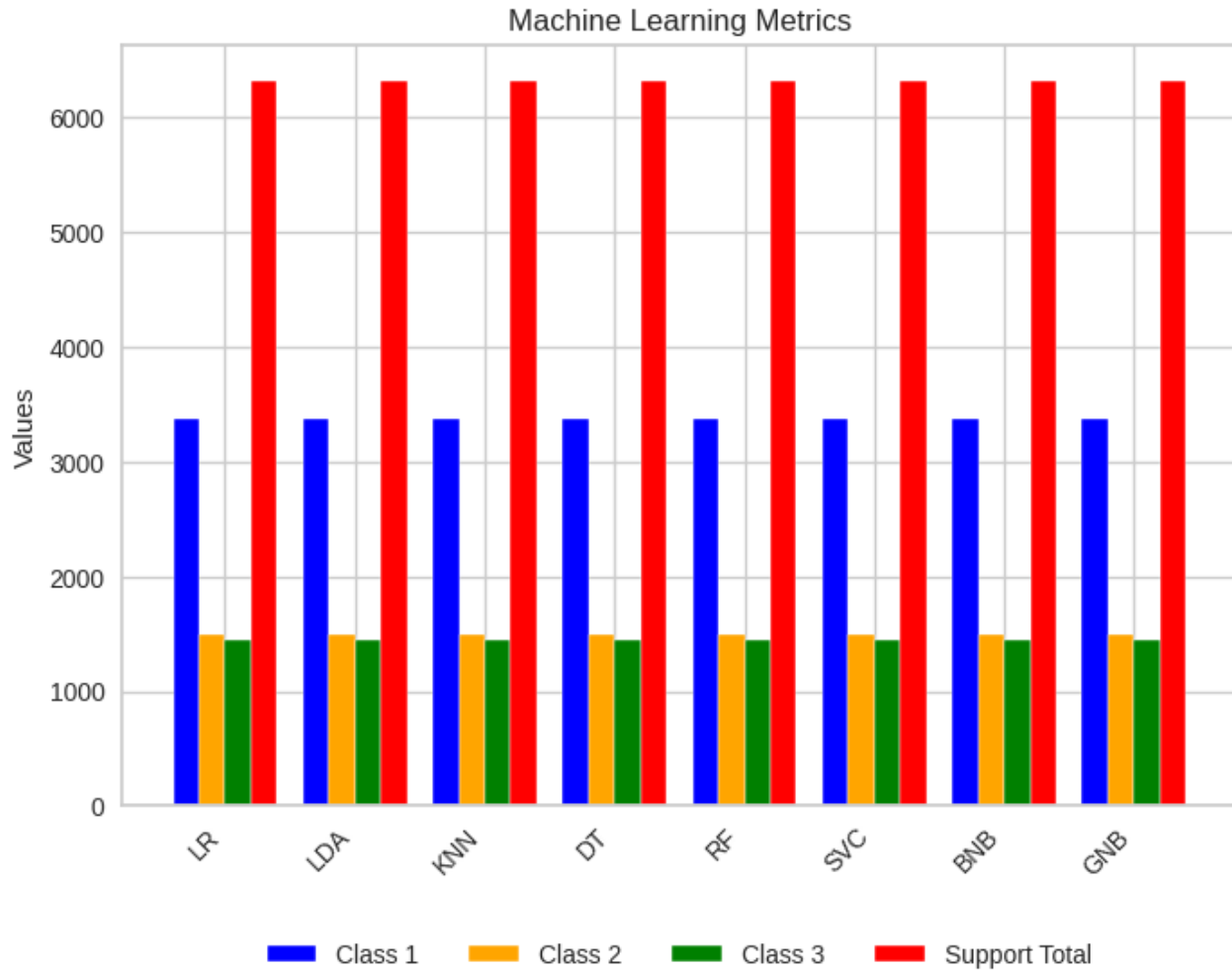

Figure 43 Support Comparison

In Figures 35-42, we did a multiclass classification ROC Curve using macro average for all of the machine learning models that we used, we were able to see that every model gave us 100% for predictions the roc curve is at a 1.0 even for the macro average amongst all the 3 classes except Linear Discriminant Analysis which gave us much poorer resulting for predictions, the value of class 0 was 0.79 which is a value that is acceptable but not perfect, anything above 0.8 to 0.9 is considered to be an excellent value for a ROC curve, but anything above 0.9 is an outstanding value, however looking at the average of all the ROC curve classes we got a score 0.97, which is considered outstanding for a ROC curve, since we are basing it off macro roc curve sampling, this model is not a good model though overall to use for predictions, since it scored low in precision, recall, f1-score, and accuracy, the ROC curve did add a good example as to why it is not good since it received a low score for class 0.

In Figure 43, we were able to see that our dataset was balanced amongst all of the classes each one of the models we trained had the same support, and the same number of actual occurrences of each class in the dataset which is a good indicator that we have a balanced dataset.

```python
from sklearn.metrics import precision_score, recall_score, f1_score, accuracy_score

# Assuming you have true labels in y_true and predicted labels in y_pred
precision = precision_score(y_test, y_pred, average='macro')
recall = recall_score(y_test, y_pred, average='macro')
f1 = f1_score(y_test, y_pred, average='macro')
accuracy = accuracy_score(y_test, y_pred)

print("Precision: {:.2f}%".format(precision*100))
print("Recall: {:.2f}%".format(recall*100))
print("F1 Score: {:.2f}%".format(f1*100))
print("Accuracy: {:.2f}%".format(accuracy*100))
```

```
Precision: 93.43%
Recall: 91.52%
F1 Score: 91.50%
Accuracy: 94.19%
```

Figure 44 Machine Learning Metrics Code

In Figure 44, you can see our machine learning metrics code, if you are interested in displaying accurate results of your machine learning model, you can print machine learning metrics to the screen such as precision, recall, f1 score and accuracy using the following code in the picture.

Figure 45 Code for ROC Curve

In Figure 45, we showed the ROC Curve code that can be used for a multiclass classification problem like this, you have to modify it to whatever machine learning model you are using it for and the specific parameters for that model and it will generate the model for you.

```python
import matplotlib.pyplot as plt
import numpy as np

# Define the data for each metric and model
class1 = [3380, 3380, 3380, 3380, 3380, 3380, 3380, 3380]
class2 = [1494, 1494, 1494, 1494, 1494, 1494, 1494, 1494]
class3 = [1443, 1443, 1443, 1443, 1443, 1443, 1443, 1443]
total = [6317,6317,6317,6317,6317,6317,6317,6317]
models = ['LR', 'LDA', 'KNN', 'DT', 'RF', 'SVC', 'BNB', 'GNB']

# Convert precision, recall, and f1_score lists to percentages
precision = [round(p, 2) for p in precision]
recall = [round(r, 2) for r in recall]
f1_score = [round(f, 2) for f in f1_score]

# Set the width of each column
width = 0.2

# Create an array of indices to center each column
x = np.arange(len(models))

# Create the figure and axes objects
fig, ax = plt.subplots()

# Create the column graph, with a different color for each metric
ax.bar(x - 1.5*width, class1, width, label='Class 1', color='blue')
ax.bar(x - 0.5*width, class2, width, label='Class 2', color='orange')
ax.bar(x + 0.5*width, class3, width, label='Class 3', color='green')
ax.bar(x + 1.5*width, total, width, label='Support Total', color='red')

# Set the labels for the x-axis and y-axis
ax.set_xticks(x)
ax.set_xticklabels(models, rotation=45, ha='right')
ax.set_ylabel('Values')

# Set the title for the graph
ax.set_title('Machine Learning Metrics')

# Add a legend to the graph
ax.legend(loc='upper center', bbox_to_anchor=(0.5, -0.15), ncol=4)

# Display the graph
plt.show()
```

Figure 46 Code for any type of column graph

In Figure 46, we used this code to generate a column graph for comparing support between models, and we modified this code to make our column graph for comparing the precision, recall, f1-score, and accuracy of our models, you can do this for any model that you would like as well, just change the values and change the names of the columns of the graph you want it to be. Overall, we were able to get the overall classification metrics for all the models except for the ANN, the ROC curve can't be printed for this model, since it's a multiclass ANN neural network, and only ROC curves can be made for a binary classification model, but based on the accuracy, precision, recall, and f1 score of that model you were able to tell that type of neural network should be avoided. It shows that you can't solely evaluate a neural network on accuracy you have to take into the account the loss as well, and also all of the other's metrics such as f1 score, precision, recall, and accuracy.

XI. More Information about Dataset, Benefits, Main Contribution of this report

The meaning of each item in our dataset is simple to understand, for the IoT, LTE, GBR, Non-GBR, AR/VR, Healthcare, industry, IoT Devices, Public Safety, Smart City, Smart Transportation, Smartphone columns, and slice type column, the most important key takeaway from this is the slice type, this is exactly what we explained in the architecture of a communications slice, type 1 is designated for the eMBB slice, type 2 is for URLLC slice and type 3 is for MMTC slice. Each row is depicted by either a 0 or a 1, 1 means that it is in use for those cases, 0 means it is not. The other columns, such a latency delay, latency loss, time, LTE category, each row depicts an amount of time it took to do a task and if there was any loss in the transmission. The most unique column was the LTE/5G column, it included set User Equipment categories that specify the performance standards. This dataset was most likely generated by a computer, or maybe through the use of crowdsourcing since there is too many items in the dataset, it is specifically not stated how this dataset was created. The benefits of this dataset to my report are that it gave us the ability to analyze communications slicing datasets, which are pretty unique there isn't many out there to look at since its fairly a new topic in the domain. This dataset allowed our Machine Learning models to predict a network slice that is appropriate each type of use case that we wanted to predict for, we were able to view the accuracy and prediction results for each model that wanted to look at individually. Another benefit of this dataset is that it wasn't too large, compared to other datasets, which means it took less time for our machine learning models to be trained against the dataset. The main contributions of this dataset to our report are that, it was preprocessed for us which made it for easy for anyone to use, it included basic metadata, which allowed the reader to understand all of the categories that were included inside of the file and what each row means by a number that was associated either a 0 or 1, where 0 means not required and 1 means its required. The dataset matched our domain that were researching on which is perfect, usually you can't find an exact dataset that will work for a report. It also included Quality statements that covered important aspects such as data accuracy, completeness, and uncertainty, allowing the end-user to make important decisions about our dataset.

Main Contributions to this report

(1) In [42] (Borgesen et al.), evaluates machine learning approaches for detecting cybersecurity threats, this is a major contribution to our report, due to the similarities in machine learning models used, we took into consideration what best models to use for our report and we choose similar models, and we picked and chose the same platform to use which is Google Collaborate, this report gave us ideas on what type of code we should work with for our machine learning models.

(2) In [43], (Kholidy et al.), creates an experimental 5G Testbed to gain the ability to test and evaluate his network slicing capabilities within his 5G Testbed. Overall, this report is also a main contribution to our report, since it talks about communications slicing in much greater detail, and evaluates the results using multiple open-source software, such as Metasploit and Snort. The security aspect is what we also covered in our report, that is the main thing that we took into consideration from his report, to use as a section for our report, and the overall talk about radio access networks (RANS), was also used as an idea from his report. The overall talk about network slicing and the three main types of slicing was also used as an idea for our report.

(3) In [44], (Kholidy et al.), talks about a Zero-Trust (ZT) model, and talks about how it can be used for secure management of communications slices, for an 5G open-architecture. It talks about overall how poor the security is currently for communications slices. This is a main contribution to our report, we used this as an idea for our report, we emphasized how important security is for communications slices and how it should be improved. The author overall did a longer approach to researching security on communications slicing.

(4) In [45], (Kholidy), talks about 5G Networks and how it is important to implement communications slicing for these networks. He uses his 5G Testbed, and uses a vulnerability tool known as Nessus, to test for vulnerabilities in the 5G network. This is also a main contribution for our report, we chose 5G because 5G is an important topic and it is the future of networking, and his talks about communications slicing are very important, we used his report as an idea to choose a topic that will be important for the future of 5G networking.

(5) In Communications slicing is an important domain in cybersecurity and the next-generation networks, it is important to learn about the architecture of communications slicing and increase security between slices in a communications network.

(6) From our report it is important to understand that for communications slicing it is important prevent denial of service attacks from occurring on a network also it is important to study and understand the application cases when you might need to use communications slicing.

(7) From our report it is important study machine learning models as they are important for prediction on many cybersecurity problems looking at problems with communications slicing you will be able to determine what ML model to use in the future if you are faced with a problem like this in the cybersecurity domain, this is something that can be used in the workplace if you have a problem you have to solve you can do it with machine learning or deep learning.

(8) From our report it is also important to know that if you want to get a job with any of the major network carriers in the USA or anywhere around the world, you must understand communications slicing if you have to work on a next-generation network.

**Acknowledgement**

I would like to thank Dr. Hisham Kholidy, for allowing me to work on the network slicing topic on 5G, this is a new topic 5G is incredible and is the future and Machine Learning is also the future! It was very interesting to learn more about Machine Learning and explore my knowledge more deeply into network slicing. I appreciate and applaud everyone in the Cybersecurity department in State University Of New York Polytechnic Institute giving me support on my work on this research on network slicing.